\documentstyle[psfig,preprint,aps,array]{revtex}

\newcommand{\T}{\Theta}
\newcommand{\Tr}{\Theta_{R}}
\newcommand{\Tt}{\Theta_{T}}
\newcommand{\ra}{\langle\hspace{.05in}}
\newcommand{\la}{\hspace{.05in}\rangle}
\newcommand{\x}{x_{o}}
\newcommand{\z}{x_{1}}
\newcommand{\1}{_{1}}
\newcommand{\2}{_{2}}

\begin{document}
\draft
\preprint{}
\title{The Best Copenhagen Tunneling Times}
\author{M. Abolhasani\footnote{E-mail: ablhasan@netware2.ipm.ac.ir}$^
{1,2}$, \& M. Golshani\footnote{E-mail: golshani@ihcs.ac.ir}$^{1,2}$}
\address{$^1$Institute for Studies in Theoretical Physics and
Mathematics (IPM), P.O.Box 19395-5531, Tehran, Iran.}
\address{$^2$Department of Physics, Sharif University of Technology,
P.O.Box 11365-9161, Tehran, Iran.}
\date{\today}
\maketitle
\begin{abstract}
Recently, people have caculated tunneling's characteristic times within 
Bohmian mechanics. Contrary to some characteristic times defined within 
the framework of the standard interpretation of quantum mechanics, 
these have reasonable values. Here, we introduce one of available 
definitions for tunnelling's characteristic times within the standard 
interpretation as the best definition that can be accepted for the 
tunneling times. We show that, due to experimental limitations, Bohmian 
mechanics leads to same tunneling times.

\end{abstract}
\section{Introduction}
A problem which does not have a clear cut answer in quantum mechanics, 
is the time that it takes for an electron to pass through a potential 
barrier. This is a problem that is important from both a theoretical
perspective\cite{cush,bed} and a technological view\cite{ran,cap}.

In quantum mechanics, time enters as a parameter rather than an 
observable (to which an operator can be assigned). Thus, there is no 
direct way to calculate tunneling times. People have tried to introduce 
quantities which have the dimension of time and can somehow be 
associated with the passage of the particle through the barrier. These 
efforts have led to the introduction of several times, some of which are 
completely unrelated to the others [5-17]. Some people have used Larmor 
precession as a clock\cite{baz} to measure the duration of tunneling for 
a steady state \cite{ryb,butt} or for a wave packet\cite{fal}. Others, 
have used Feynman paths like real paths to calculate an average tunneling 
time with the weighting function $exp[iS(x(t))/\hbar]$, where $S$ is the
action associated with the path $x(t)$- where $x(t)$'s are Feynman paths 
initiated from a point on the left of the barrier and ending at another 
point on the right of it\cite{sok}. On the other hand, a group of people 
have used some features of an incident wave packet and the comparable 
features of the transmitted packet to introduce a delay as tunneling 
time\cite{hauge1,hauge2}. There are many other approaches, some of which 
are mentioned in Refs. [10-17]. But, there is no general consensus among 
physicists about the meaning of them and about which, if any, of them 
being the proper tunneling time. In Bohmian mechanics\cite{bohm}, 
however, there is a unique way of identifying the time of passage 
through a barrier. This time has a reasonable behaviour with respect to 
the width of the barrier and the energy of particle\cite{l1,l2}. 

It is expected that with the availability of reliable experimental 
results in the near future, an appropriate definition can be selected 
from the available ones, or that they would prepare the ground for a 
more appropriate definition of the transmission time. But now, we want 
to use the definition of tunneling time in the framework of Bohmian 
mechanics to select one of available definitions for quantum tunneling 
times (QTT) within the standard interpretation as the best definition. 

Our paper is organized as follows: after introducing Olkhovsky-Recami 
QTT, by using a heuristic argument in section II, we introduce, in 
section III, Bohmian QTT. Then, in section IV, we give a critical 
discussion about Cushing's thought experiment and about what it really 
measures.  

\section{Tunneling's characteristic times in the Copenhagen framework}
To begin with, we consider the time at which a particle passes through 
a definite point in space. We describe the particle by a Gaussian wave 
packet which is incident from the left. The most natural way to estimate 
this time of passage is to find the time at which the peak of the wave 
packet passes through that point. But this is not a right criterion for 
finding the time of passage of the particle (even if the wave packet is
symmetrical). To clarify the matter, we divide the packet, in the 
middle, into two parts. The probability of finding the particle in the 
front section is $\frac{1}{2}$ and the same is true for the back 
section. We represent the transit time of the centre of gravity of the 
front section by $t_{1}$ and that of the back section by $t_{2}$. The 
average time for particle's passage through that point is $T=\frac{1}
{2}(t_{1}+t_{2})$. If the transit time for the peak of the wave is 
denoted by $t$, we have: 
\begin{mathletters}     
\begin{eqnarray}
t_{1}=t-\frac{x_{1}}{v_{g}}
\end{eqnarray}
\begin{eqnarray}
t_{2}=t+\frac{x_{2}}{v_{g}}
\end{eqnarray}
\end{mathletters}
where $v_{g}$ is the group velocity of the wave packet and $x_{1}$ and 
$x_{2}$ are, respectively, the distances of the centers of gravity of 
the front and the back sections of the packet from its peak position, 
when these centers pass the point under consideration. Thus, we have:

\begin{eqnarray}
T=\frac{1}{2}(t_{1}+t_{2})= t+\frac{x_{2}-x_{1}}{2v_{g}}
\end{eqnarray}

If the wave packet did not spread, $x_{1}$ and $x_{2}$ would remain
equal and $T$ would be equal to $t$. But, since the wave packet spreads, 
$T\ne t$. In fact, the average transit time for the particle is later 
than that of wave's peak. Because, the spreading of the packet decreases 
the transit time of the centre of gravity of wave's front section, and 
increases that of the back 
\begin{figure}
\centerline{\psfig{figure=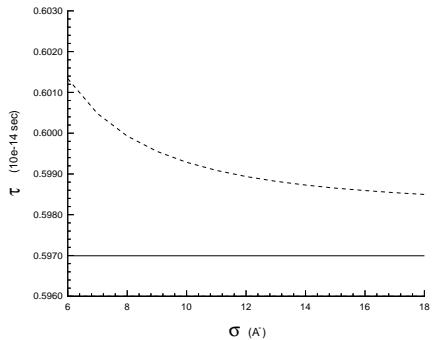,width=.4\columnwidth}}
\caption{{\footnotesize Average time for the transmission of the 
probability flux ($---$) and the passage of the peak (------), for
packets having different widths ($\sigma$).}
}
\end{figure}

section. But the change is not symmetrical 
(i.e. $x_{1}\ne x_{2}$), as the back section of the wave experiences the 
spreading for a longer time.

Now, consider a wave packet $\psi(x,t)$, which is incident from the left 
and approaches a far point $x$. The best time that we can attribute to 
particle's passage through $x$ is 

\begin{eqnarray}
\tau(x)=\frac{\int_{0}^{\infty}t\ |\psi(x,t)|^2\  v(x,t)\ dt}{\int_
{0}^{\infty}\ |\psi(x,t)|^2\ v(x,t)\ dt}
\end{eqnarray}
where $v(x,t)=${\Large$\frac{j(x,t)}{|\psi(x,t)|^2}$}, $j(x,t)$ being 
the probability current density. In fact, we have divided the wave 
packet into infinitesimal elements. The transit time when particle is 
in one of these elements is weighted by the probability of finding the 
particle there (i.e. $|\psi(x,t)|^2\ dx=|\psi(x,t)|^2\ v(x,t)\ dt$). 
Fig.(1) illustrates the difference between this time and the time that 
the peak passes that point. For narrow wave packets, for which the rate 
of spreading is large, this difference is large. From (3), one can 
define a distribution for the transit time through $x$:

\begin{eqnarray}
P(x,t)=\frac{|\psi(x,t)|^2\ v(x,t)}{\int_{0}^{\infty} |\psi(x,t)|^2
\ v(x,t) dt}=\frac{j(x,t)}{|T|^2}
\end{eqnarray}
where $|T|^2$ is the transition probability for passing through $\x$. 
Dumont and Marchioro introduced this definition for the distribution of 
the time at which a particle passes through the far side of a potential 
barrier\cite{dum}. They did not find it possible  to define the time 
spent by the particle in the barrier. Leavens showed that this is also 
the distribution for the same time in Bohmian mechanics\cite{l3}.

By looking at (3), one notices that $\tau(x)$ is in fact the average 
time for the passage of the probability density $|\psi|^2$ through $\x$. 
Since the probability density represents the probability of the presence 
of the particle, it is natural to take the average time for the passage 
of probability density through a point as a measure of the average time
for particle's passage through that point. But, while part of the 
probability flux passes through the barrier, the particle itself might not 
be detected on the other side of the barrier. We don't, however, expect 
to get a definite prediction for an individual system, and in the 
laboratory we usually consider an ensemble of systems. Thus, it is 
natural to take the average time for the passage of the probability 
density as a measure of the average time for particle's passage. From 
now on, we talk about particle's average time of transit. Consider, a 
particle incident on a barrier from the left. Then, one can easily 
extend (3) to define average times for particle's entrance into the 
barrier ($\tau_{_{in}}$), particle's exit from the right side of the 
barrier ($\tau_{out}^{^{T}}$), and particle's exit from the left side 
of the barrier ($\tau_{out}^{^{R}}$). To simplify the matter we use the
following notations:

\begin{eqnarray}
(\ \ldots\T\ )_{x}^{\pm}=\int_{0}^{\infty}\ dt\ \ldots\ (\pm)j(x,t)\ \T
[\pm j(x,t)]
\end{eqnarray}
where $j(x,t)$ represents the probability current density at the point 
$x$ at time $t$, and $\T$ is the usual step function. Using this 
definition and (3), we define $\tau_{_{in}}$, $\tau_{_{out}}^{^R}$ and 
$\tau_{_{out}}^{^T}$ as:
\begin{mathletters}
\begin{eqnarray}
\tau_{_{in}}=\frac{\int_{0}^{\infty}\ dt\ t\ j(a,t)\ \T[+j(a,t)]}{\int
_{0}^{\infty}\ dt\ j(a,t)\ \T[+j(a,t)]}=\frac{(\ t\ \T\ )_{a}^{^+}}{(\ 
\T\ )_{a}^{^+}}
\vspace{5mm}
\\
\tau_{_{out}}^{^{T}}=\frac{\int_{0}^{\infty}\ dt\ t\ j(b,t)\ \T[+j(b,t)
]}{\int_{0}^{\infty}\ dt\ j(b,t)\ \T[+j(b,t)]}=\frac{(\ t\ \T\ )_{b}^{^
+}}{(\ \T\ )_{b}^{^+}}
\vspace{5mm}
\\
\tau_{_{out}}^{^{R}}=\frac{\int_{0}^{\infty}\ dt\ t\ (-)j(a,t)\ \T[-
j(a,t)]}{\int_{0}^{\infty}\ dt\ (-)j(a,t)\ \T[-j(a,t)]}=\frac{(\ t\ 
\T\ )_{a}^{^-}}{(\ \T\ )_{a}^{^-}}
\end{eqnarray}
\end{mathletters}
where $a$ and $b$ represent the coordinates of the left and right side
of the  barrier respectively. Using these times, one can write, the
times that particle spends in the barrier before the transmission ($\tau_
{_{T}}^{^{OR}}$) or reflection ($\tau_{_{R}}^{^{OR}}$) as:
\begin{mathletters}
\begin{eqnarray}
\tau_{_{T}}^{^{OR}}=\tau_{_{out}}^{^T}-\tau_{_{in}}
\\
\tau_{_{R}}^{^{OR}}=\tau_{_{out}}^{^R}-\tau_{_{in}}
\end{eqnarray}
\end{mathletters}
We shall call them OR times\footnote{Note that, in their orginal 
definition, temporal integrations run from $-\infty$ to $+\infty$. In 
Ref \cite{orz}, they discussed that the substitution of integrals of 
the type $\int^{\infty}_{0}$ for integrals $\int_{-\infty}^{+\infty}$ 
have physical significance. In any way, we shall use relations (6).} 
(referring to Olkhovsky and Recami\cite{olr}). The average time spent 
by the particle in the barrier, irrespective of being transmitted or 
reflected, the so called dwelling time, is thus given by 
\begin{eqnarray}
\tau_{_{d}}^{^{OR}}=(\ \T\ )_{b}^{^+}\tau_{_{T}}^{^{OR}}+(\ \T\ )_{a}^
{^-}\tau_{_{R}}^{^{OR}}
\end{eqnarray}
where $(\T)^{^+}_{b}$ and $(\T)^{^-}_{a}$ represent the probability of 
particle's exit from the right and left sides of the barrier 
respectively. Now, the probability of particle's exit from the right, 
$(\T)^{^+}_{b}$, is equal to the probability of particle's transmission 
through the barrier, $|T|^2$. But the probability of particle's exit 
from the left, $(\T)^{^-}_{a}$, is not equal to the probability of 
reflection from the barrier, $|R|^2$. Because, the particle could be 
reflected without entering the barrier. Using (7), one can write (8) in 
the form:
\begin{eqnarray}
\tau_{_{d}}^{^{OR}}=(\ \T\ )_{b}^{^+}\ \tau_{_{out}}^{^{T}}+(\ \T\ )_{a}
^{^-}\ \tau_{_{out}}^{^{R}}-(\ \T\ )_{a}^{^+}\tau_{_{in}}
\end{eqnarray}
where we have made use of the fact that $(\ \T\ )_{b}^{^+}+(\ \T\ )_{a}
^{^-}=(\ \T\ )_{a}^{^+}$, which follows from the conservation of 
probability. The first two terms in (9) represent the average of 
particle's exit time from the barrier, irrespective of the direction of 
exit. Using (6) we can write the right hand side of (9) in the form:

\begin{eqnarray}
\tau_{_{d}}^{^{OR}}=\int_{0}^{\infty}\ dt\ t\ [j(b,t)-j(a,t)]
\end{eqnarray}
Using continuty equation, one can easily show that (10) coincides with
the standard dwelling time defined by

\begin{eqnarray}
\tau_{_{D}}=\int_{0}^{\infty}\ dt\ \int_{a}^{b}|\psi(x,t)|^2\ dx
\end{eqnarray}

\section{Tunneling's characteristic times in Bohmian framework}

In the causal interpretation of quantum mechanics, proposed by David 
Bohm\cite{bohm}, a particle has a well defined position and velocity 
at each instant, where the latter is obtained from a field $\psi(x,t)$ 
satisfying the Schr\"odinger equation. If the particle is at $x$ at 
the time $t$, its velocity is given by

\begin{eqnarray}
v(x,t)=\frac{j(x,t)}{|\psi(x,t)|^2}
\end{eqnarray}
For a particle which is prepared in the state $\psi(x,0)$ at $t=0$, any 
uncertainty in its dynamical variables is a result of our ignorance 
about its initial position $\x$. Our information about particle's 
initial position is given by a probability distribution $|\psi(\x,0)|
^2$. If we know the initial position $\x$ of the particle, we can find 
its position at a later time, $x(\x;t)$, from (12). Then, when a 
particle encounters a barrier, it is determined whether the particle 
passes through the barrier or not, and one can determine when the 
particle enters the barrier and when it leaves the barrier. Thus the 
time spent by the particle within the barrier is easily calculated. But, 
since we do not know particle's initial position, we consider an 
ensemble of initial positions, given by the distribution $|\psi(\x,t)|
^2$. Then, we calculate the average time spent by the particle within the 
barrier. To compare the time of reflection or transmission in this 
framework with OR charactristic times, we first consider the time of 
arrival at $\z$, for a particle that was at $\x$ at $t=0$

\begin{eqnarray}
t(\z;\x)=\int_{C_{\x}}dx\ t(x;\x)\ \delta(\z-x)
\end{eqnarray}
where the integral is defined along Bohmian path $C_{\x}$ which 
starts at $\x$. This relation can also be written in the form:

\begin{eqnarray}
t(\z;\x)=\int_{0}^{\infty}\ dt\ |v(x(\x;t),t)|\ t\ \delta(\z-x(\x;t))
\end{eqnarray}
where

\begin{eqnarray}
\delta(\z-x(\x;t)) =\frac{\delta(t(\z)-t)}{|v(x(\x;t),t)|}
\end{eqnarray}
Since it is possible for the particle to pass the point $\z$ twice (due 
to reflection from the barrier), we define $t^{\pm}(\x;\z)$ in the 
following manner:

\begin{eqnarray}
t^{\pm}(\x;\z)=\int_{0}^{\infty}\ dt\ |v(x(\x;t),t)|\ t
\ \delta(\z-x(\x;t))\ \T[\pm v(x(\x;t),t)]
\end{eqnarray}
where $t^{+}$ and $t^{-}$ correspond to the cases where the particle 
passes $\z$ from left to right and from right to left respectively. 
Since for long periods of time, a particle either passes or is 
reflected (depending on its $\x$), we define $\Tr$ and $\Tt$ in the 
following way\cite{l1,l2}:

\begin{mathletters}
\begin{eqnarray}
\Tt(\x)=1,\ \ \Tr(\x)=0 \hspace{.3in} (for\ transmission)
\\
\Tt(\x)=0,\ \ \Tr(\x)=1 \hspace{.53in} (for\ reflection)
\end{eqnarray}
\end{mathletters}
Thus, we have $\Tt(\x)+\Tr(\x)=1$. Using these functions, the 
average times spent by the transmitted and the reflected particles, $
\tau_{_{T}}^{^B}$ and $\tau_{_{R}}^{^B}$, respectively, are given by

\begin{mathletters}
\begin{eqnarray}
\tau_{_{T}}^{^{B}}=\frac{\ra t^{+}(\x;b)\ \Tt(\x)\la-\ra t^{+}(\x;a)
\ \Tt(\x)\la}{\ra \Tt(\x)\la}
\\
\tau_{_{R}}^{^{B}}=\frac{\ra t^{-}(\x;a)\ \Tr(\x)\la-\ra t^{+}(\x;a)
\ \Tr(\x)\la}{\ra \Tr(\x)\la}
\end{eqnarray}
\end{mathletters}
where

\begin{eqnarray}
\ra\ldots\la=\int_{-\infty}^{+\infty}\ d\x\ \ldots\ |\psi(\x,t)|^2
\end{eqnarray}
But $\ra \Tr(\x)\la=|T|^2$ and $\ra\Tr(\x)\la=|R|^2$\cite{l1}. Thus, we 
have for the dwelling time:

\begin{eqnarray}
\tau_{d}^{^{B}}=|T|^2\tau_{_{T}}^{^{B}} +|R|^2\tau_{_{R}}^{^{B}}\hspace
{1.26in}
\nonumber
\\
=\ra t^{+}(\x;b)\ \Tt(\x)\la+\ra t^{-}(\x;a)\ \Tr(\x)\la-\ra t^{+}
(\x;a)\la
\end{eqnarray}
where we have made use of the fact that $\Tt+\Tr=1$. Using the fact 
that $\int_{-\infty}^{+\infty}d\x\ f(x(\x,t),t)\ |\psi(\x,0)|^2\ \delta
(x-x(\x,t))=f(x,t)\ |\psi(x,0)|^2$, one can easily show that

\begin{mathletters}
\begin{eqnarray}
\ra t^{+}(\x;b)\ \Tt(\x)\la=(\ t\ \T\ )_{b}^{^+}
\\
\ra t^{-}(\x;a)\ \Tr(\x)\la=(\ t\ \T\ )_{a}^{^-}
\\
\ra t^{+}(\x;a)\la=(\ t\ \T\ )_{a}^{^+}
\end{eqnarray}
\end{mathletters}
Thus, $\tau_{_{d}}^{^B}$ is equal to $\tau_{_{d}}^{^{OR}}$ and therefore 
equal to $\tau_{_{D}}$. Of course, the equality of $\tau_{_{d}}^{^{B}}$ 
and $\tau_{_{D}}$ was shown earlier by Leavens\cite{l1,l2}. But the 
relations (21) are new and they are important because they show the 
relation between OR characteristic times and those defined in Bohmian 
mechanics. Notice that in the causal interpretation of Bohm, one defines 
two average entrance times $\tau_{in}^{^T}$ and $\tau_{in}^{^R}$, 
depending on wether the particle is reflected or transmitted:

\begin{mathletters}
\begin{eqnarray}
\tau_{in}^{^T}=\frac{\ra t^{+}(\x,a)
\ \Tt(\x)\la}{\ra \Tt(\x)\la}
\\
\tau_{in}^{^R}=\frac{\ra t^{+}(\x,a)
\ \Tr(\x)\la}{\ra \Tr(\x)\la}
\end{eqnarray}
\end{mathletters}
whereas OR have defined only one average time. This is because in the
standard interpretation of quantum mechanics, it is not definite whether 
a particle that has entered a barrier, is transmitted or reflected. It 
is natural to have the average time for particle's entrance, 
irrespective of wether it is reflected or transmitted, to be equal to $
|T|^2\tau_{in}^{^T}+|R|^2\tau_{in}^{^R}=\ra t^{+}(\x;a) \la$ in 
Bohmian framework. Then, we must have $(\ t\ \T\ )_{a}^{^+}=|T|^2\tau_
{in}^{^T}+|R|^2\tau_{in}^{^R}$, which is easy to prove. 

It is natural to expect that the average time for particle's transmision 
through a potential barrier to be a function of the width of the 
barrier. This time should generally increase with the width of barrier. 
However, due to quantum effects, one does not expect it to be a linear 
function of this width. Most of the times defined within the framework 
of the standard interpretation of quantum mechanics, do not have this 
property, and some of them even yield negative times! On the other hand, 
we expect the transition time to decrease with the increase in the 
energy of the incident particle. The transition time in OR approach has 
both of these properties . The digrams in Fig.(2) and Fig.(3) represent 
the trasmission time as a function 
of the width of the barrier and as a function of particle's energy 
respectively, for both Bohmian and OR times. The numerical method used 
to solve the time 
dependent Schr\"odinger equation was the fourth order (in time steps $
\delta t$) symmetrized product formula method, developed by De Readt
\cite{der}. We chose $\delta x=${\Large$\pi$}$/30k_{o}$, where  $k_{o}=
\sqrt{\frac{2mE_{o}}{\hbar^2}}$ and $\delta t={\delta x}^2/25$ in all 
calulation ($E_{o}$ is the energy of the incident Gaussian wave packet). 
One notices that OR transmission time coincides with that of Bohmian 
case for large $|T|^2$ (i.e. $d<2$ in the diagrams of Fig(2) and $E_{o}
>V_{o}$ in diagram of Fig(3)). This is natural, because while the 
average time for particle's exit from the right side of the barrier is 
always the same in both approaches, in the limit of $|T|^2\rightarrow 
1$, the average entrance time for the tansmitted particle is the same 
in OR approach and in Bohmian approach ($|T|^2\rightarrow 1 \Rightarrow  
\tau_{_{in}}^{^T}\rightarrow \tau_{_{in}}$). Thus in this limit we have 
$\tau_{_{T}}^{^B}=\tau_{_{T}}^{^{OR}}$. On other hand, as we said 
earlier, the average time for particle's exit from the left, in OR 
approach, is generally different from that of Bohmian case. But, if we 
choose $a$ to be a point far (relative to the width of wave packet) from 
the left side of the barrier, then, the time for particle's exit from 
the left side is the same in both approaches. Since, in this case for 
$|R|^2\rightarrow 1$, the average time of entrance for reflected 
particles in causal approach become equal to the average time of 
entrance in OR approach ($|R|^2\rightarrow 1 \Rightarrow  \tau_{_{in}}
^{^R}\rightarrow \tau_{_{in}}$). Thus we have $\tau_{_{R}}^{^B}=\tau_
{_{R}}^{^{OR}}$. It appears that OR approach gives the most natural 
definition for a positive definite transmission time, within the 
framework of the standard interpretation of quantum mechanics\footnote
{Of course, in relation (6b) if we chose $b$ a point in the interior of 
barrier width, $\tau_{_{T}}^{^{OR}}$ maybe become nagative, but small in 
absolute value\cite{dbm,l4}}. 
\begin{figure}
\centerline{\psfig{figure=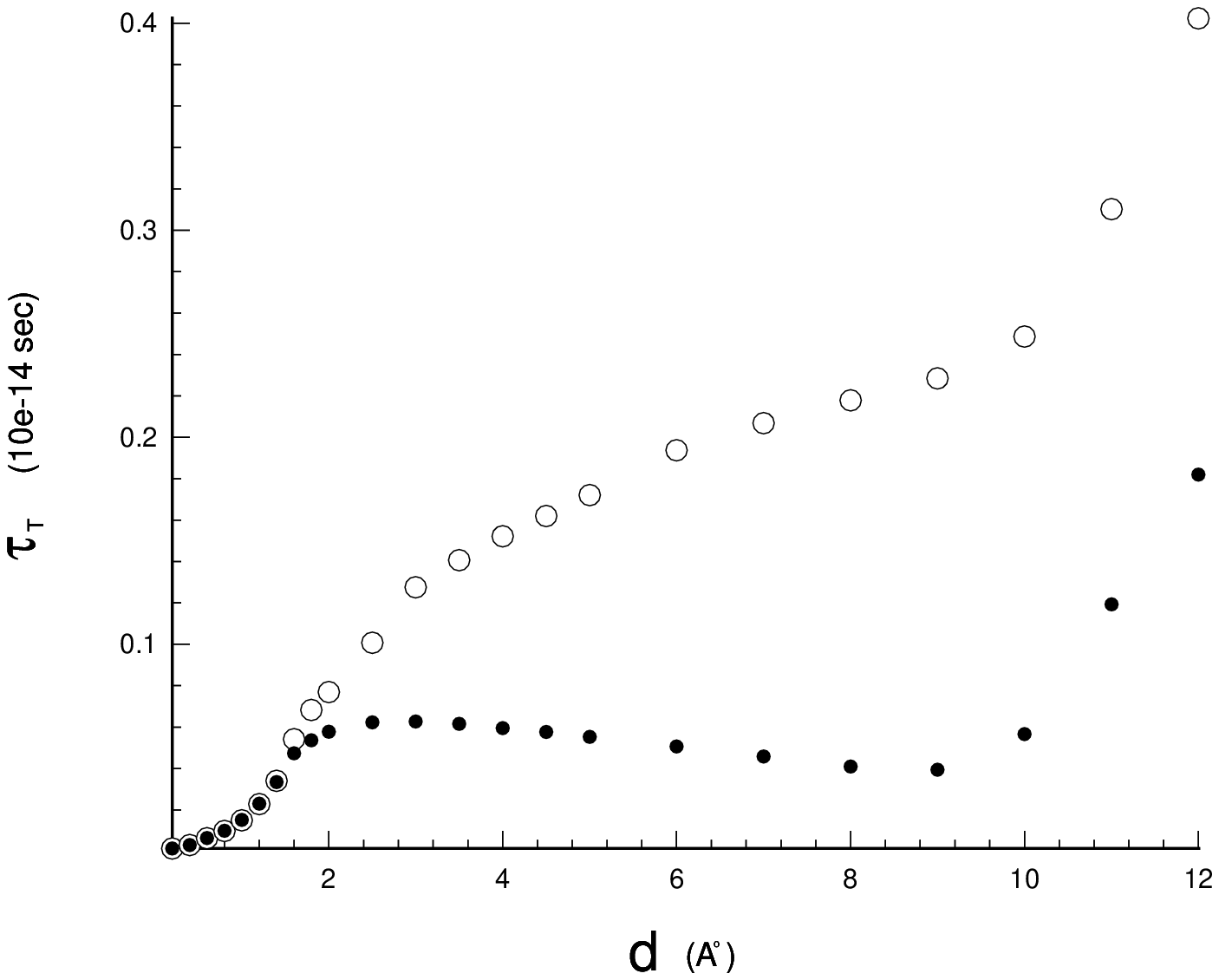,width=.4\columnwidth}}
\centerline{(a)}
\end{figure}
\begin{figure}
\centerline{\psfig{figure=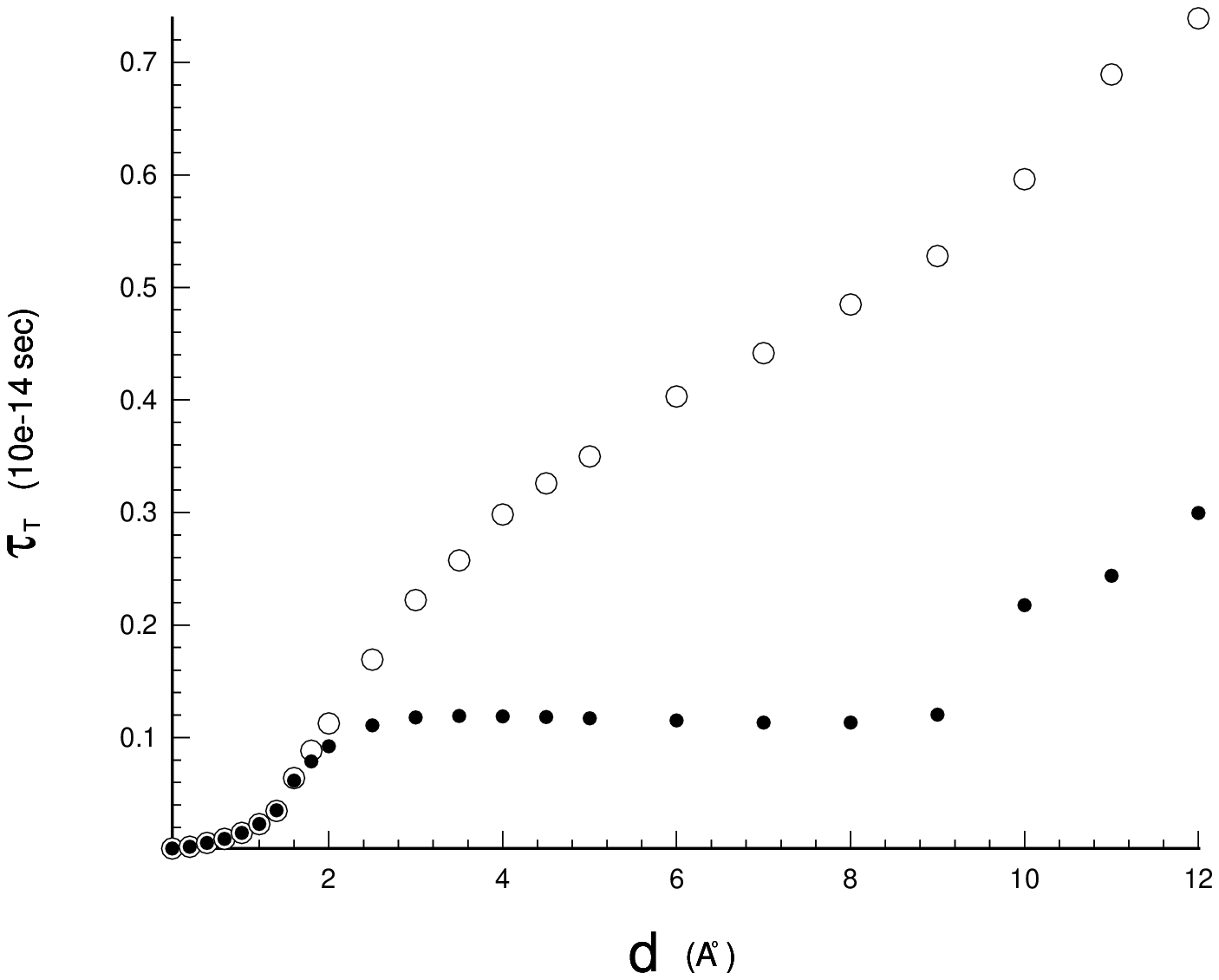,width=.4\columnwidth}}
\centerline{(b)}
\end{figure}
\begin{figure}
\centerline{\psfig{figure=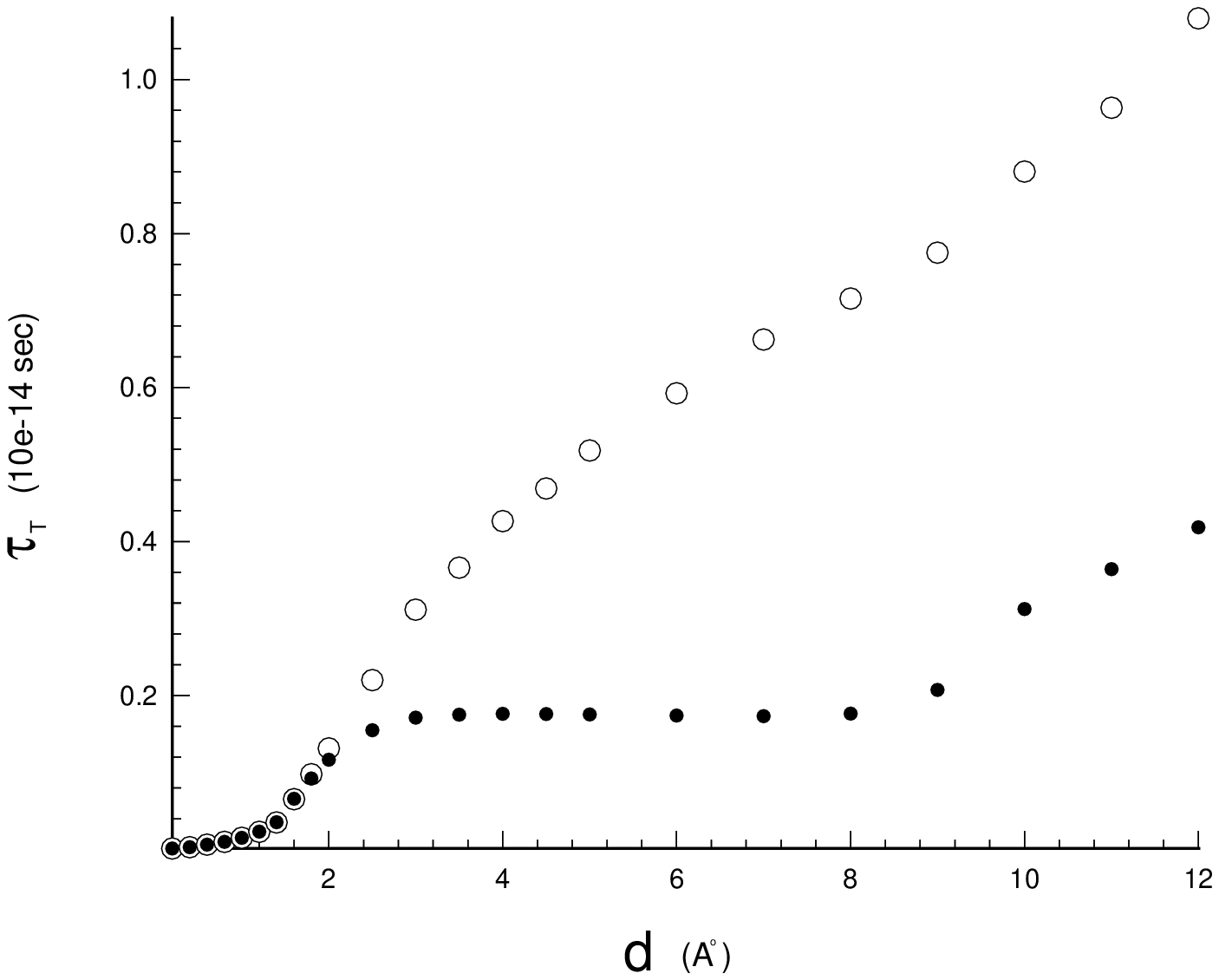,width=.4\columnwidth}}
\centerline{(c)}
\vspace{5mm}
\caption{{\footnotesize Diagrams (a), (b) and (c) show the dependence of 
the transmission time $\tau_{_{T}}$ in terms of the width of a square 
barrier with the height $V_{o}=10 eV$ and the incident energy $E_{o}=
\frac{\hbar^2 k_{o}^2}{2m}=5eV$. These diagrams represent, respectively, 
Gaussian wave packets having the width $\sigma=6 A^{o}, \sigma=12 A^{o}$ 
and $\sigma=18 A^{o}$. We have shown the Bohmian results by hollow 
spheres and those of the standard interpretation by solid circles.}
}
\end{figure}

\begin{figure}
\centerline{\psfig{figure=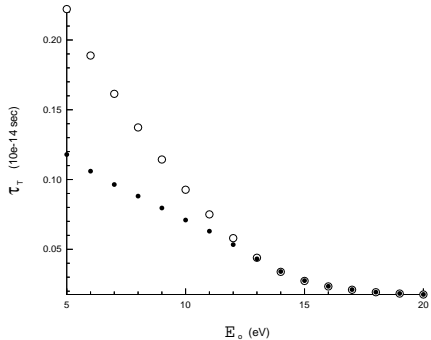,width=.4\columnwidth}}
\caption{{\footnotesize This diagram shows the transit time for a 
square barrier of width $b-a=3 A^{o}$ and of height $V_{o}=10 eV$, for 
Gaussian wave packets having $\sigma=12 A^{o}$ and differents energies. 
We have shown Bohmian results by hollow spheres and those of the 
standard interpretation by solid circles.}
}
\end{figure}

\section{Experimental test}
It is generally believed that the standard quantum mechanics and Bohmian
mechanics have identical predictions for physical observables. On the 
other hand, there is no Hermitian operator associated with time. Is it 
possible to consider a phenomenon involving time, e.g. tunneling, to 
differentiate between these two theories? By considering a thought 
experiment, Cushing gave a positive response to this question. His 
argument was the following \cite{cush,bed}:\\
(1) There is presently no satisfactory account of a quantum tunneling 
time (QTT) in the standard quantum mechanics.\\
(2) There is a well defined account of QTT in Bohmian interpretation. If 
it can be measured, then such a measurement would constitute a test of 
the interpretation .\\
(3) It might be possible to measure the Bohmian QTT with an experiment of 
a certain type.\\
(4) Therefore, from (2) and (3), if an experiment of that type is 
possible, such an experiment could serve as a test of Bohm's 
interpretation.\\
(5) Because of (1), the outcome of an experiment of that type would not 
support or refute the copenhagen interpretation.\\
In a recent article, K. Bedard\cite{bed}, by refering to (1), (2) and 
(5) questioned Cushing's conclusion. Her argument was based on the fact 
that the two theories have different microontologies. Therefore, the 
QTT obtainable from Bohmian mechanics has no counterpart in the standard 
quantum mechanics. Thus, the measurement of such a time cannot be 
considered a test between the two theories. Here, we shall question (3), 
i.e. the claim that Cushing's thought experiment can be used to measure 
Bohmian times.

Cushing's experiment consists of a potential barrier between $x\1$ and 
$x\2$ with width $d$ ($d=b-a$). A detector $D_{T}$ is located at $x\2$ 
on the right of barrier and a detector $D_{R}$ is located at $x\1$ on 
the left of the barrier ($x\1<a<b<x\2$). Electrons are incident from 
the left. $D_{T}$ records the arrival times of the transmitted electrons 
at $x_{2}$ and $D_{R}$ records the times of the reflected electron 
at $x\1$. The distance from $x\1$ to the left side of the barrier ($a$) 
is a much more than width of wave packet. The same holds for $x\2$. 
The recording of the arrival time of the incident electron at $x\1$ 
will collapse the wave function. In that case, any subsequent tunneling 
time prediction on the basic of the known incident wave packet would be 
quite useless. To resolve this problem, Cushing considers the 
preparation of the state of the incident particle at $x\1$, rather than 
its detection. Thus, the time recorded at $x\1$ is the preparation time 
for the transit of the particle, if $D_{T}$ would detect it, and the 
preparation time for the reflected paticle, if $D_{R}$ would detect it. 
To provide this condition, we prepare a source of electrons in front of 
which there is a shutter. The shutter starts to open little before 
$t_{o}=0$ and closes little after $t_{o}$.  Thus $t_{o}$ is the most 
probable time for the passage of the electron from $x\1$. In other 
words, $t_{o}$ is the time when the peak of the wave packet passes 
$x\1$. By choosing a weak source, we can be sure to have at most one 
electron emerging from shutter's opening. The time of passage for the 
particle through $x\1$ is $t\1=t_{o}\pm\frac{\Delta x}{v_{o}}$, where $
\Delta x$ is the widths of the packet and $v_{o}$ is the speed of the 
particle.  Cushing claims that " in principle, this error could be made 
as small as we like (for large enough $v_{o}$)"\cite{cush}. In our 
opinion, the error must be campared with $\tau_{_{T}}$ not with $t\1$. 
In fact, we want to obtain $\tau_{_{T}}$ which is the difference of the 
two time ($t_{2}-t\1$), where $t_{2}$ is the time that electron is 
detected at $x_{2}$ by $D_{T}$). The error could be small if we compare 
it with $t\1$ and $t_{2}$ but not if we compare it with their 
difference. Thus, we must have:

\begin{eqnarray}
\frac{\Delta x}{v_{o}}\ll \tau_{_{T}}
\end{eqnarray} 
By refering to the Fig.(3), one can see that $\tau_{_{T}}$ decreases 
quicker than $\frac{1}{v_{o}}$. Thus, the increase in $v_{o}$ decreases 
the right hand side of (23) more than its left hand side and we are not
able to decrease relative error in this way. One may hope obtaining 
condition (23) by decreasing $\Delta x$. But decreasing $\Delta x$ is 
not useful. Because, by refering to Fig.(2) (a, b, c) one can see that 
$\tau_{_T}$ decreases almost linearly with $\Delta x$.

Experimental limitations dictate that the arrival time of particles to 
the barrier be measured independent of whether they shall be reflected 
or transmitted (i.e. state preparation time). Thus, although Bohm's 
theory considers $\tau_{_{in}}^{^T}$ and $\tau_{_{in}}^{^R}$, it must 
pay attention only to the measurements of $\tau_{_{in}}=|T|^2\tau_{_
{in}}^{^T}+|R|^2\tau_{_{in}}^{^R}$, to avoid experimental limitations. 
In this way, the precise time that it attributes to particle's 
transmission through the barrier or reflection is $\tau_{_{T}}^{^{OR}}$ 
and $\tau_{_{R}}^{^{OR}}$ respectively. Thus, at the experimental level, 
even in the case of tunneling times we have the same predictions in the 
two theories. In fact, here, we encounter a problem like the case of the 
celebrated two-slit experiment. In the framework of Bohmian mechanics, 
all particles observed on the lower (upper) half of the screen must come 
from the lower (upper) slit. But, any effort to know which particle came 
from which slit destroys the interference pattern. Thus, in the two-slit 
experiment, the two theory come to the same result due to experimental 
limitations. It appears that, from various definitions given for QTT in 
the framework of the standard quantum mechanics, our choice of OR's is 
the best. Because, in our opinion it is the best time that can be 
related to the tunneling phenomena in the framework of the standard 
interpretation of quantum mechanics. We can justify our claim in the 
follow way:\\
(1) There is a unique and well defined account of QTT in Bohm's 
interpretation.\\
(2) There are several accounts of QTT in standard interpretation.\\ 
(3) These two theories have the same prediction for observables.\\
(4) Bohmian prediction for QTT coincides with one of Copenhagen QTT 
(OR's).\\

In fact, OR's is the only definition that gives the same result, at the 
experimental level, as Bohmian mechanics, although it does not associate 
an operator with $\tau_{_{T}}$ (at least up to now). In this way, we 
have used a theory with additional microontology (Bohmian mechanics) to 
give the best definition for a quantity in a theory with less 
microontology. Bohm's theory may also shed light on other definitions of 
QTT in the standard quantum mechanics.

\centerline{\bf Conclusion}
Considering the fact that the microontology of Copenhagen theory 
includes wave function (probability amplitude), and not point-like 
particles, the best time one could attribute to the passage of a 
particle from a point of space is the average time of the passage 
of probability flux (eq.(3)). Generalization of this time to QTT, 
leads one to OR's times. On other hand, the microontology of Bohmian 
mechanics includes point-like particles in addition to wave function, 
and it leads uniquely to Bohmian QTT (eq.(18)). We have compared them 
for different width and energy of wave packet in Fig.(2) and Fig.(3) by 
use of numerical calculation. 

Now, Bohmian QTT could not be measured due to experimental limitations.
The best times that could be obtained in Bohmian mechanics are the same 
as OR's. The agreement of one of the several\footnote{In fact, only 
systematic projector approach of Brouard, Sala and Muga leads to an 
infinite hierarchy of possible mean transmission and reflection times
\cite{bsm}.} available definitions of QTT in Copenhagen quantum 
mechanics with the unique definition of Bohmian mechanics, separates it 
from others. Because, it is reasonable to expect same prediction for the 
two theories even in the case of QTT.

\centerline{\bf Acknowledgments}
The authors would like to thank Dr. C. R. Leavens drawing our attention 
to prior work of Olkhovsky and Recami.

\end{document}